\documentclass[runningheads,a4paper]{llncs}

\makeatletter
\DeclareRobustCommand*\textsubscript[1]{%
  \@textsubscript{\selectfont#1}}
\def\@textsubscript#1{%
  {\m@th\ensuremath{_{\mbox{\fontsize\sf@size\z@#1}}}}}
\makeatother

\usepackage{amssymb}
\usepackage{amsmath} 
\usepackage{rotating}
\usepackage{float}
\usepackage{pgfplots} 
\usepackage{xcolor}
\usepackage{subcaption}
\usepackage{float}
\usepackage{pgfplots} 
\usepackage{tikz}
\usepackage{tikzscale}
\usepackage{graphicx}
\usepackage{algorithm}
\usepackage{bbm}
\usepackage{setspace}
\usepackage[noend]{algpseudocode}
\usepackage{url}

\urldef{\mailshash}\path|{shashank.gupta}@research.iiit.ac.in|
\urldef{\mailprofs}\path|{manish.gupta, vv}@iiit.ac.in|
\urldef{\mailtcs}\path|{sachin7.p, nitin.ramrakhiyani,gk.palshikar}@tcs.com|
\newcommand{\keywords}[1]{\par\addvspace\baselineskip
\noindent\keywordname\enspace\ignorespaces#1}
\definecolor{bblue}{HTML}{4F81BD}
\definecolor{rred}{HTML}{C0504D}
\definecolor{ggreen}{HTML}{9BBB59}
\definecolor{ppurple}{HTML}{9F4C7C}

\renewcommand\textfloatsep{4pt}

\renewcommand\intextsep{4pt}

\begin{document}

\setlength{\abovedisplayskip}{3pt}
\setlength{\belowdisplayskip}{3pt}
\mainmatter  

\title{Co-training for Extraction of Adverse Drug Reaction Mentions from Tweets}


%
%
\author{Shashank Gupta\inst{1} \and Manish Gupta\inst{1}\thanks{Author is also a Principal Applied Scientist at Microsoft} \and 
Vasudeva Varma\inst{1} \and Sachin Pawar\inst{2} \and Nitin Ramrakhiyani\inst{2} \and Girish Keshav Palshikar\inst{2}}
\authorrunning{Gupta et al.}
\institute{International Institute of Information Technology-Hyderabad, India\\
\email{shashank.gupta@research.iiit.ac.in}\\ \email{\{manish.gupta,vv\}@iiit.ac.in}
\and
TCS Research, Pune\\
\email{\{sachin7.p,nitin.ramrakhiyani,gk.palshikar\}@tcs.com}}

%
%

\maketitle

\begin{abstract}
Adverse drug reactions (ADRs) are one of the leading causes of mortality in health care. Current ADR surveillance systems are often associated with a substantial time lag before such events are officially published. On the other hand, online social media such as Twitter contain information about ADR events in real-time, much before any official reporting. Current state-of-the-art methods in ADR mention extraction use Recurrent Neural Networks (RNN), which typically need large labeled corpora. Towards this end, we propose a semi-supervised method based on co-training which can exploit a large pool of unlabeled tweets to augment the limited supervised training data, and as a result enhance the performance. Experiments with $\sim$0.1M tweets show that the proposed approach outperforms the state-of-the-art methods for the ADR mention extraction task by $\sim$5\% in terms of F1 score. 

\end{abstract}

\section{Introduction}
Estimates show that Adverse Drug Reactions (ADRs) are the fourth leading cause of deaths in the United States ahead of cardiac diseases, diabetes, AIDS and other fatal diseases\footnote{https://ethics.harvard.edu/blog/new-prescription-drugs-major-health-risk-few-offsetting-advantages}. 
Hence, it necessitates the monitoring and detection of such adverse events to minimize the potential health risks.
Typically, post-marketing drug safety surveillance (also called as pharmacovigilance) is conducted to identify ADRs after a drug's release. Such surveys rely on formal reporting systems such as Federal Drug Administration's Adverse Event Reporting System (FAERS)\footnote{http://bit.ly/2xnu7pE}. However, often a large fraction ($\sim$94\%) of the actual ADR instances are under-reported in such systems~\cite{hazell2005under}. Social media presents a plausible alternative to such systems, given its wide userbase. A recent study~\cite{freifeld2014digital} shows that Twitter has three times more ADRs reported as compared to FAERS. 

Earlier work in this direction focused on feature based pipeline followed by a sequence classifier \cite{nikfarjam2015pharmacovigilance}. More recent works are based on Deep Neural Networks \cite{cocos2017deep,DBLP:conf/eacl/StanovskyGM17}. Deep learning based methods~\cite{collobert2011natural,lecun2015deep} typically rely on the presence of a large annotated corpora, due to their large number of free parameters. Due to the high cost associated with tagging ADR mentions in a social media post and limited availability of labeled datasets, it is hard to train a deep neural network effectively for such a task. In this work, we attempt to address this problem and propose a novel semi-supervised method based on co-training \cite{blum1998combining} which can harness a large pool of unlabeled related tweets, which are more economical to collect than ADR annotated tweets.   

\section{Approach}\label{approach}
In this section, we define the ADR extraction problem, discuss the supervised ADR extraction method, and then propose our semi-supervised co-training method.
\subsection{Problem Definition} 

The problem of ADR extraction can be defined as follows. Given a social media post in the form of a word sequence $x=x_1. ..., x_n$, where $n$ is the maximum sequence length, predict an output sequence $y_1,....,y_n$, where each $y_i$ is encoded using standard sequence labeling encoding scheme such as the IO encoding similar to that used in~\cite{cocos2017deep}.

\subsection{Supervised ADR Extraction} We choose the model described in~\cite{cocos2017deep} for modeling the ADR extraction task. Given an input word sequence $x$, a bi-directional LSTM transducer (bi-LSTM) \cite{DBLP:journals/corr/abs-1211-3711} is employed to capture complex sequential dependencies. Formally, at each time-step $t$, the bi-LSTM transducer attempts to model the task as follows.

\begin{equation}
h_t = \text{bi-LSTM}(e_t, h_{t-1}) 
\end{equation}
where $h_t \in \mathcal{R}^{(2\times d_h)}$, is the hidden unit representation of the bi-LSTM with $d_h$ being the hidden unit size. Since it is a concatenation of hidden units of a forward sequence LSTM and a backward sequence LSTM, its overall dimension is $2\times d_h$. $e_t$ is the embedding vector corresponding to the input word $x_t$ extracted from a pre-trained word embedding lookup table.
\begin{eqnarray}
y_t = \text{softmax}(W\times h_t + b) 
\end{eqnarray}
\noindent where $y_t \in \mathcal{R}^{d_l}$, is the output vector at each time-step which encodes the probability distribution over the number of possible output labels ($d_l$) at each time-step of the sequence. $W \in \mathcal{R}^{{d_l*d_h}}$ and $b \in \mathcal{R}^{d_l}$ are weight vectors for the affine transformation. Finally, the cross entropy loss function for the task is defined as follows.
\begin{eqnarray}
L_{\text{ADR}} = - \sum_{t=1}^{n} \sum_{i=1}^{d_l}  \hat{y_{t_i}} \log y_{t_i}\label{adr-loss}
\end{eqnarray}
\noindent where $\hat{y_t}$ is the one-hot representation of the actual label at time-step $t$.

\subsection{Co-training Method for ADR Extraction}
Algo. \ref{algo1} outlines the method for semi-supervised co-training. Co-training \cite{blum1998combining} requires two feature views of the dataset, which in our case are: (1) word2vec embeddings trained on a generic tweet corpus \cite{godin2015multimedia} followed by a bi-LSTM feature extractor and (2) word2vec embeddings trained on domain specific (drug-related) tweet corpus (described in the Section~\ref{sec:expts}) followed by a bidirectional Gated Recurrent Unit (bi-GRU) transducer~\cite{cho2014properties}. At each step of co-training, the transducers $M_1$ and $M_2$ are trained on their respective views minimizing the ADR training loss (Line \ref{step4} to \ref{step5} of Algo. \ref{algo1}). Each sample from the unlabeled example pool is scored using a scoring function computed as follows. First, the current transducer is used to decode/infer output label distribution for each word in the unlabeled sample. For each word in the output sequence, we simply choose the output label which has the maximum probability. We filter out the data sample if the transducer does not output even a single ADR label for any word in the sample. If there is at least one word labeled as ADR, we compute the score for the sample as the multiplication of the ADR probabilities for the ADR-labeled words in the sample normalized by the number of ADR words. If this confidence score of the sample is greater than some pre-defined threshold $\tau$, the sample is added to the training set of the other transducer along with its output labels as generated by the transducer (Lines~\ref{algo2:line6} to~\ref{algo2:line9}).  
Due to this cross-exchange of training data, both transducers work in synergy and learn from mistakes of each other. 

\begin{algorithm}[t!]
\scriptsize
\caption{\scriptsize Co-training Method for ADR Extraction}\label{algo1}
\hspace*{\algorithmicindent} \textbf{Input} \hspace{0.3mm}  $U$: Large collection of unlabeled tweets, $\tau$ : Threshold for co-training\\
\hspace*{\algorithmicindent} \hspace*{\algorithmicindent} \hspace*{\algorithmicindent}   $D_{ADR}^1$, $D_{ADR}^2$  : Two views of the ADR annotated data \\
\hspace*{\algorithmicindent} \textbf{Output} \hspace{0.3mm} Model parameters $\theta^{LSTM}$, $\theta^{GRU}$
\begin{algorithmic}[1]
\State $T^1$, $T^2$ $\gets$ $D_{ADR}^1$, $D_{ADR}^2$
\State Initialize model parameters, $\theta^{LSTM}$, $\theta^{GRU}$ randomly.
\While{\textit{(stopping criteria is not met)}}
\State $M^{1}$ $\gets$ train bi-LSTM on $T^{1}$ (minimize $L_{ADR}^1$)\label{step4}
\State $M^{2}$ $\gets$ train bi-GRU on $T^{2}$ (minimize $L_{ADR}^2$)\label{step5}
\For{$i\gets 1, \vert U \vert$}
\If{$M^1.score$($U_i$) $ \ge$ $\tau$ }\label{algo2:line6}
\State $T^2 \gets T^2 \cup \{U_i\}$, $U\gets U - U_i$\label{algo2:line7}
\EndIf
\If{$M^2.score$($U_i$) $ \ge$ $\tau$ }\label{algo2:line8}
\State $T^1 \gets T^1 \cup \{U_i\}$, $U\gets U - U_i$\label{algo2:line9}
\EndIf
\EndFor
\EndWhile \label{algo2:step2}
\end{algorithmic}
\end{algorithm}

\section{Experiments}
\label{sec:expts}
In this section, we discuss details of the datasets, implementation details and experimentation results.
\subsection{Datasets}
We use two datasets for evaluation detailed as follows. 

\noindent\underline{\textbf{(1) Twitter ADR}}: The \textit{Twitter ADR} dataset, described in \cite{cocos2017deep}, contains 957 tweets posted between 2007 and 2010, with mention annotations of ADR and some other medical entities. Due to Twitter's license agreement, authors released only tweet ids with their corresponding mention span annotations. At the time of collection of the original tweets using Twitter API, we were able to collect only 639 tweets with 1526 ADR mentions. 

\noindent\underline{\textbf{(2) TwiMed}}: The \textit{TwiMed} dataset, described in \cite{alvaro2017twimed}, contains 1000 tweets with mention annotation of Symptoms from drug (ADR) and other mention annotations posted in 2015. Due to Twitter's license agreement, we were able to extract 663 tweets only with 1091 ADR mentions. 

\noindent\underline{\textbf{Unlabeled Tweets}}: For semi-supervised learning, we collected $\sim$0.1M tweets using the keywords as drug-names and ADR lexicon publicly available\footnote{http://diego.asu.edu/downloads}. This filtering step ensures that all collected tweets have at least one drug-name occurrence and one ADR phrase. The tweets were posted in 2015 and have no ADR mentions labeled.


\subsection{Implementation Details}
For implementation of the model, we use the popular Python deep learning toolkits: Keras\footnote{https://keras.io/} and TensorFlow\footnote{https://www.tensorflow.org/}. Training data for each fold is divided according to 90:10\% train-validation split. 

\noindent\underline{\textbf{Pre-processing}} As part of text pre-processing, all HTML links and user mentions are normalized to a single token respectively. Special characters and emoticons are removed, and each tweet is padded with the maximum tweet length in the corpus. 

\noindent\underline{\textbf{Hyper-parameter settings for the two views:}} The hyper-parameter settings for the two views as required by the co-training method are as follows. 

\noindent\underline{\textbf{View 1}}: For the first view we use bi-LSTM transducer, with the hyper-parameter setting similar to the one reported in \cite{cocos2017deep}. Word embedding dimension is set to 400.  

\noindent\underline{\textbf{View 2}}: For the second view, we use bi-GRU transducer with input as word2vec word embeddings trained on the unlabeled drug-related tweets described in the previous section. The word embedding dimension is set to 300.

For both transducers, the hidden unit dimension ($d_h$) is set to 500. The number of output units ($d_l$) is 4. We use Adam \cite{kingma2014adam} as optimizer with a learning rate of 0.001 and a batch size of 32. 

\noindent\underline{\textbf{Co-training Parameters}}: For the co-training methods, confidence threshold value is empirically set to 0.5. The stopping criteria for the co-training kicks in when the number of iterations reaches 5 or if the unlabeled tweets pool is exhausted, whichever occurs first.  The number of epochs are set to a maximum with 25, with early-stopping employed if validation loss drops for more than 3 epochs.

\subsection{Results}

\begin{table}
\centering
\resizebox{\textwidth}{!}{%
 \begin{tabular}{|l | c | c | c |c | c | c| } 
 \hline
 \textbf{Method} & \multicolumn{3}{c|}{\textbf{Twitter ADR Dataset}} & \multicolumn{3}{c|}{\textbf{TwiMed Dataset}} \\
 \hline
& \textbf{Precision} & \textbf{Recall} & \textbf{F1-score} & \textbf{Precision} & \textbf{Recall} & \textbf{F1-score} \\
 \hline
 Baseline \cite{cocos2017deep} &  0.7067$\pm$0.057 & 0.7207$\pm$0.074 & 0.7102$\pm$0.049 &  0.6120$\pm$ 0.116 &		0.5149$\pm$ 0.099 &		0.5601$\pm$ 0.100 \\
 Baseline with Adam &  0.7065$\pm$0.058 & 0.7576$\pm$0.083 &	0.7272$\pm$0.051&  \textbf{0.6281$\pm$0.094}	&	0.5614$\pm$0.110	&	0.5859$\pm$0.079  \\
 KB-Embedding  & 0.7171$\pm$ 0.058 &	0.7713$\pm$0.091 &	0.7397$\pm$0.055 & 0.5960$\pm$0.081 &		0.6144$\pm$0.068	&	0.6042$\pm$0.060 \\
Baseline \cite{DBLP:conf/eacl/StanovskyGM17}&&&&&&\\
 \hline 
  Co-training (5k) & 0.7247$\pm$0.056 &	0.7770$\pm$0.082 &	0.7488$\pm$ 0.063 & 0.5806$\pm$0.093 &	0.6746$\pm$0.078 &	\textbf{0.6192$\pm$0.066}\\
  Co-training (10k) & 0.7288$\pm$0.041 &	\textbf{0.8238$\pm$0.064} &	0.7719$\pm$0.040 & 0.5484$\pm$0.092	&	0.6355$\pm$0.113	&	0.5851$\pm$0.090\\
  Co-training (25k) & 0.7181$\pm$0.035 &	0.8005$\pm$0.048	& 0.7561$\pm$0.031 & 0.5774$\pm$0.082	&	0.6425$\pm$0.076	&	0.6051$\pm$0.066 \\
  Co-training (50k) & 0.7207$\pm$0.034 &	0.7870$\pm$0.042 &	0.7516$\pm$0.029 & 0.5420$\pm$0.054	&	0.6342$\pm$0.061	&	0.5836$\pm$0.053\\
  Co-training (75k) & 0.7478$\pm$0.062 &	0.8033$\pm$0.053 &	0.7730$\pm$0.047 & 0.5525$\pm$0.059	&	\textbf{0.6875$\pm$0.069}	&	0.6110$\pm$0.056\\
  Co-training (100k) & \textbf{0.7514$\pm$0.053}	& 0.8045$\pm$0.056 &	\textbf{0.7754$\pm$0.042} & 0.5548$\pm$0.064	&	0.6786$\pm$0.058	&	0.6081$\pm$0.048 \\
  \hline
\end{tabular}
}
\caption{Accuracy Comparison for Various Methods (along with Std. Deviation)}\label{tab:resultsComb}
\end{table}

The results of various methods are presented in Table~\ref{tab:resultsComb} for the Twitter ADR  and the TwiMed datasets respectively. For the ADR task, to encode the output labels we use the IO encoding scheme where each word is labeled with one of the following labels: (1) I-ADR (ADR mention), (2) I-Other (mention category other than ADR), (3) O (others), (4) PAD (padding token). Since our entity of interest is ADR, we report the results on ADR only. An example tweet annotated with IO-encoding is as follows. ``\textit{@BLENDOS\textsubscript{O} Lamictal\textsubscript{O} and\textsubscript{O} trileptal\textsubscript{O} and\textsubscript{O} seroquel\textsubscript{O} of\textsubscript{O} course\textsubscript{O} the\textsubscript{O} seroquel\textsubscript{O} I\textsubscript{O} take\textsubscript{O} in\textsubscript{O} severe\textsubscript{O} situations\textsubscript{O} because\textsubscript{O} weight\textsubscript{I-ADR} gain\textsubscript{I-ADR} is\textsubscript{O} not\textsubscript{O}  cool\textsubscript{O}}''. For performance evaluation, we use approximate-matching \cite{tsai2006various}, which is used popularly in biomedical entity extraction tasks \cite{cocos2017deep,nikfarjam2015pharmacovigilance}. We report the F1-score, Precision and Recall computed using approximate matching as follows.
\small

\begin{equation}
\text{Precision} = \frac{\text{\#ADR approx. matched}}{\text{\#ADR spans predicted}}, \text{Recall}= \frac{\text{\#ADR approx. matched}}{\text{\#ADR spans in total}}
\end{equation}

\normalsize

The F1-score is the harmonic-mean of the Precision and Recall values. All results are reported using 10-fold cross-validation along with the standard deviation across the folds. Our baseline methods are bi-LSTM transducer \cite{cocos2017deep} with traditional word embeddings and the current state-of-the-art bi-LSTM transducer which uses traditional word embeddings augmented with knowledge-graph based embeddings \cite{DBLP:conf/eacl/StanovskyGM17}. For both the datasets, it should be noted that Cocos et al. \cite{cocos2017deep} used RMSProp as an optimizer, and since we are using Adam for all our methods, so for a fair comparison we also report the baseline results with Adam. The corresponding results are reported in the first two rows of Table~\ref{tab:resultsComb}. It is clear that re-implementation with Adam optimizer enhances the performance, which is consistent with the general consensus around Adam optimizer. The KB-embedding baseline \cite{DBLP:conf/eacl/StanovskyGM17} replaces word embeddings of the medical entities in the sentence with the corresponding embeddings learned from a knowledge-base. The corresponding results can be seen in row 3. It is clear that adding KB-based embeddings enhances the performance over the baseline, due to the external knowledge added in the form of KB embeddings. 

The results for our methods are presented from row 4 onwards. It is clear that the co-training method outperforms the baseline by a significant margin. It clearly indicates the efficacy of semi-supervised learning when the labeled data is scarce. 

\noindent \underline{\textbf{Effect of Unlabeled Data Size:}} We also analyze the effect of the size of the unlabeled tweet dataset on the method's performance. The results are presented from row 4 onwards. The results are fairly constant as unlabeled data size is varied, indicating the robustness of the method.

\section{Conclusions}\label{conclusion}
In this paper, we proposed a semi-supervised co-training based learning based methods to tackle the problem of labeled data scarcity for adverse drug reaction mention extraction task. Our method uses large unlabeled drug related tweets to augment the limited existing ADR extraction datasets providing superior results in comparison to pure supervised learning based methods. We analyzed the method on two popular ADR extraction datasets, and it demonstrates superior results as compared to the state-of-the-art methods in ADR extraction. 

\end{document}